\documentclass[preprint,aps,pre,epsf, superscriptaddress]{revtex4-2}
\usepackage{amsmath}
\usepackage{amssymb}
\usepackage{epsfig}
\graphicspath {{./figures/}}
\makeatletter
\def\input@path{{./figures/}}
\makeatother

\begin{document}

\title{Compensation in continuous symmetric trilayered planar ferrimagnet}

\author{Olivia Mallick}
\email{mallickolivia0@gmail.com}
\affiliation{Department of Physics of Complex Systems, S. N. Bose National Centre for Basic Sciences, Kolkata-700106, India}

\author{Muktish Acharyya}
\email{muktish.physics@presiuniv.ac.in}
\affiliation{Department of Physics, Presidency University,
86/1 College Street, Kolkata-700073, India}

\date{\today}

\begin{abstract}
 The trilayered (A-B-A type) anisotropic (single site) XY magnetic model is considered with the intra-plane ferromagnetic and 
 the inter-plane antiferromagnetic interactions. The equilibrium properties of such trilayered anisotropic XY system has been studied by 
 Monte Carlo simulation with  Metropolis single spin (randomly chosen) update algorithm. In a given range of interaction parameters, the system has been found to exhibit the high temperature {\it critical} (vanishing sublattice magnetisations and total magnetisation) and low temperature {\it compensation} (vanishing total magnetisation even for non-vanishing sublattice magnetisations) behaviours. The compensation temperature and the critical temperature both are found to increase with the increase of single site anisotropy. The comprehensive phase diagram is drawn. The finite size study reveals the growth of correlations near the critical temperature.

\vskip 3cm



\end{abstract}

\maketitle
\newpage

\section{Introduction} The trilayered magnetic system exhibits compensation across a range of appropriately adjusted interaction parameters. The total magnetisation can be reduced to zero by tuning the temperature. At high temperature, the sublattice magnetisation in each layer vanishes, resulting in zero total magnetization. However, a notable phenomenon occurs when nonzero sublattice magnetisation collectively yield a vanishing total magnetisation, a phenomenon referred as {\it compensation}. This effect demonstrates a precise balance among the sublattice magnetisations that causes the total magnetisation to disappear.

The compensation in trilayered ferrimagnetic models has widely been studied over the last decade. Mainly, the discrete symmetric spin models (Ising, Blume-Capel, Clock model etc.) are used to study such compensation phenomena with magneticaloric effects.

Experimentally, ${\rm Mn_{1.5}FeV_{0.5}Al}$ compound showed ferrimagnetic compensation observed by magnetic and anomalous Hall
measurements\cite{expt1}. The compensated ferrimagnetism was found and the compensation temperature was estimated\cite{expt2} from the Hall resistivity in Heusler alloys. Nearly compensated ferrimagnetic behaviours and giant exchange bias have
been observed\cite{expt3} in ${\rm Mn_{2}PtAl}$. 
Recently, the complete magnetisation compensation was found in ferromagnetic semiconductor ${\rm Gd_{x}Sm_{1-x}N}$\cite{expt4}.

The simplest discrete symmetric spin model, the Ising model (1925)\cite{editorial} has usually been exploited to study the magnetic behaviours. The compensation in ferrimagnetic trilayered Ising system has been studied\cite{branco} by Monte Carlo (MC) simulation. The effects of nonmagnetic impurities on the compensation temperature in Ising trilayred model has been studied
\cite{ma1} by MC simulation. The compensation in trilayered anisotropic Blume-Capel model has been investigated\cite{ma2} to 
study the anisotropy dependency of the compensation temperature. Recently, the discrete symmetric 6-state clock model has been
used to study the compensation behaviour. The MC results showed\cite{olivia} interesting anisotropy dependent compensation temperature. The trilayered discrete symmetric mixed spin models are found\cite{hadey,dely, anti1,anti2, extra1,extra2,extra3,extra4} to show the interesting compensation behaviours. The mixed spin quantum Heisenberg models are also found to show the compensation behaviour. The phase
transition and compensation has been studied\cite{bobak} in mixed spin anisotropic Heisenberg system (described by Oguchi pair Hamiltonian). 

The Graphene bilayer has been studied\cite{comp-bilayer} by MC simulation to show the behaviours of magnetic phase transition and compensation. Reentrant spin orientation and compensation are observed\cite{re-comp} in Ising ferrimagnet. The effects of inclined magnetic field on the spin orientations and compensation have also been studied\cite{comp-field} theoretically in ferrimagnetic iron garnet with uniaxial anisotropy. The magnetocaloric effect and the critical behaviour associated with the loss of long-range magnetic order have been investigated\cite{asym} in an asymmetric ferrimagnetic bilayer using MC simulations.
 
Although, the SO(2) continuous symmetric ferromagnetic XY model shows\cite{KT} interesting and unusual kind of phase transition
(Kosterlitz-Thouless transition), in two dimensions, the trilayered ferrimagnetic XY system has not yet been used to explore the compensation behaviours. In this article, we took the initiative (may be the first time) to study the compensation in trilayered anisotropic XY model by extensive Monte Carlo simulations. The manuscript is arranged as follows: the next section (section-II)
introduces the model and  the Monte Carlo methodology in continuous symmetric planar magnetic (XY) model, the numerical results are given in section-III and the paper ends with a comprehensive summary in section-IV.

\label{sec:introduction}

\section{Model and Simulation scheme:}
\label{sec:model}
We consider a classical anisotropic XY spin model defined on a trilayered system (A-B-A) composed of stacked square lattices. Each lattice site (i) hosts a two-component unit vector spin $(\vec{S}_i = (S_i^x, S_i^y) = (\cos\theta_i,\sin\theta_i))$, where $(\theta_i \in [0,2\pi])$. The Hamiltonian of the system reads as,

\begin{equation}
    \mathcal{H}
= - \sum_{\langle i j \rangle} J_{ij}  ~ \vec{S}_i \cdot \vec{S}_j-D \sum_i \left[(S_i^x)^2 - (S_i^y)^2\right]
\label{Hamiltonian}
\end{equation}

   where the first term represents the nearest-neighbour exchange interaction and the second term corresponds to a single-site in-plane anisotropy. The summation $\langle ij\rangle$runs over all nearest-neighbour pairs of the trilayered system, including both intra- and interlayer interactions. The exchange coupling $J_{ij}$ depends explicitly on the layer indices of the interacting spins and is defined as,
  \begin{equation*}
      J_{ij} =
  \begin{cases}
  J_{\mathrm{AA}}, & i,j \in A\\
  J_{\mathrm{BB}}, & i,j \in B\\
  J_{\mathrm{AB}}, & i \in A, B; j \in B, A
  \end{cases}
  \end{equation*}

  where $(J_{\mathrm{AA}} > 0)$ and $(J_{\mathrm{BB}} > 0)$ denote ferromagnetic intralayer interactions within the outer (A) layers and the middle (B) layer, respectively, while $(J_{\mathrm{AB}} < 0)$ corresponds to the antiferromagnetic interlayer coupling between adjacent (A) and (B) layers. The anisotropy parameter $(D)$ controls the preferential alignment of spins within the $XY$ plane: for $(D>0)$, spins tend to align along the $x$-direction, whereas for $(D<0)$, alignment along the $y$-direction is favoured in order to achieve the energy minimized configurations. The competition between ferromagnetic intralayer and antiferromagnetic interlayer interactions, combined with the in-plane single site anisotropy, usually give rise to rich magnetic behaviour, i.e., layer-dependent ordering, and possible {\it compensation} phenomena in the trilayered geometry. We imposed the periodic boundary
conditions (PBC) in the x and y directions within each layer and along the stacking direction z, we adopt free boundary conditions to capture the finite thickness of the trilayered structure.

  To study the compensation behaviour in the trilayered anisotropic XY model, we employ Monte Carlo simulations based on the Metropolis algorithm with random single-spin updates. The system consists of three stacked square lattices (A–B–A), each of size $L\times L$, giving a total of $3L^2$ classical spins.  The interaction strengths $J_{AA}$, $J_{AB}$ and the anisotropy $D$ are measured in the unit of $J_{BB}$. 
  
  The intralayer exchange interactions are chosen to be ferromagnetic with unequal coupling strengths, $J_{AA}=0.5$ and $J_{BB}=1.0$, representing the different ferromagnetic environments of the constituent layers. The interlayer exchange interaction is taken to be antiferromagnetic, $J_{AB}=-0.1$. The competition between the ferromagnetic intralayer and antiferromagnetic interlayer couplings gives rise to ferrimagnetic ordering and enables the emergence of compensation behavior.The simulations are initialized at high temperature with random spin orientations uniformly distributed over the interval $[0,2\pi]$, corresponding to a disordered paramagnetic state. At each Monte Carlo step, a lattice site is selected randomly (random updating scheme) and a trial spin orientation is generated by choosing a new angle $\theta_i' \in [0,2\pi]$. The energy change $\delta E$ associated with this update is computed from Equation-\ref{Hamiltonian}, and the trial move is accepted with the Metropolis probability\cite{binder} $\mathbf{P = {\rm Min}[{1, \exp(-\delta E / k_B T)}]}$, where $k_B$ is the Boltzmann constant and $T$ is the temperature (measured in the unit of ${{J_{BB}} \over {k_B}}$). Note that the time unit in our simulations is one Monte Carlo step per site (MCSS), which consists of $3L^2$ such random updates. For each temperature, the system is allowed to equilibriate over $2\times 10^5$ MCSS, followed by an additional $3\times 10^5$ MCSS for statistical averaging (assuming the ergodic limit for the equality of the time averaging and the ensemble averaging). 

To characterize the magnetic properties, we compute layerwise magnetisation components of the sublattice magnetisations). For each layer $(k=1,2,3)$ the magnetisation components are defined as, 

\begin{equation*}
    m_k^x=\frac{1}{L^2}\sum S^x, ~~~~~~~~~~~~~~~~  m_k^y=\frac{1}{L^2}\sum S^y, 
\end{equation*}
where the summations are carried over all lattice sites in the k-th layer only. The sublattice magnetisation ($m$) of any particular sublattice
(k-th sublattice) can
be calculated as $m=\sqrt{(m_k^x)^2+(m_k^y)^2}$.

The components of the total magnetisations are defined as:

\begin{equation*}
M_x = \frac{1}{3}\sum_k m_k^x, ~~~~~~~~~~~~~~~~~~M_y = \frac{1}{3}\sum_k m_k^y,
\end{equation*} 
where the summations are carried over three ($k=1,2,3$) layers.

  The total magnetisation of the system is obtained by summing over all three layers  $M=\sqrt{(M_x)^2+(M_y)^2}$.

The corresponding magnetic susceptibility is defined by, 
\begin{equation*}
    \chi=\frac{3L^2}{k_BT}(\langle M^2 \rangle - \langle M \rangle ^2). 
\end{equation*}

The specific heat $C$ is defined as,

\begin{equation*}
    C=\frac{dE}{dT}
\end{equation*}
where $E$ is the average energy density calculated from Hamiltonian(Equation-\ref{Hamiltonian}).

\section{ Results:}
\label{sec:results}
\begin{figure}[htbp]
    \centering
	(a)\includegraphics[width=0.5\textwidth]{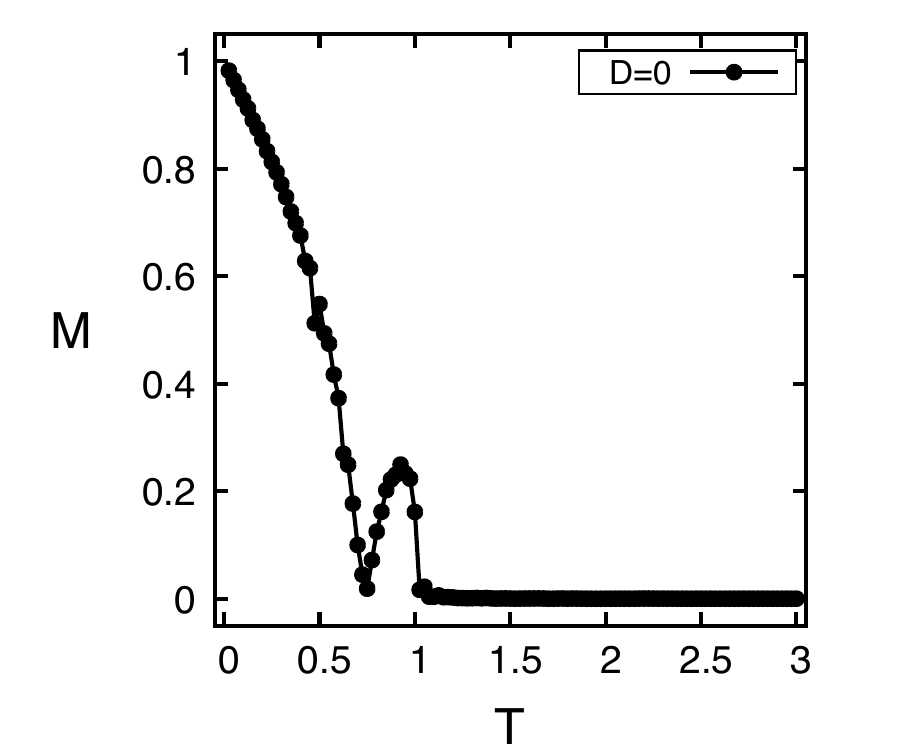}
	(b)\includegraphics[width=0.45\textwidth]{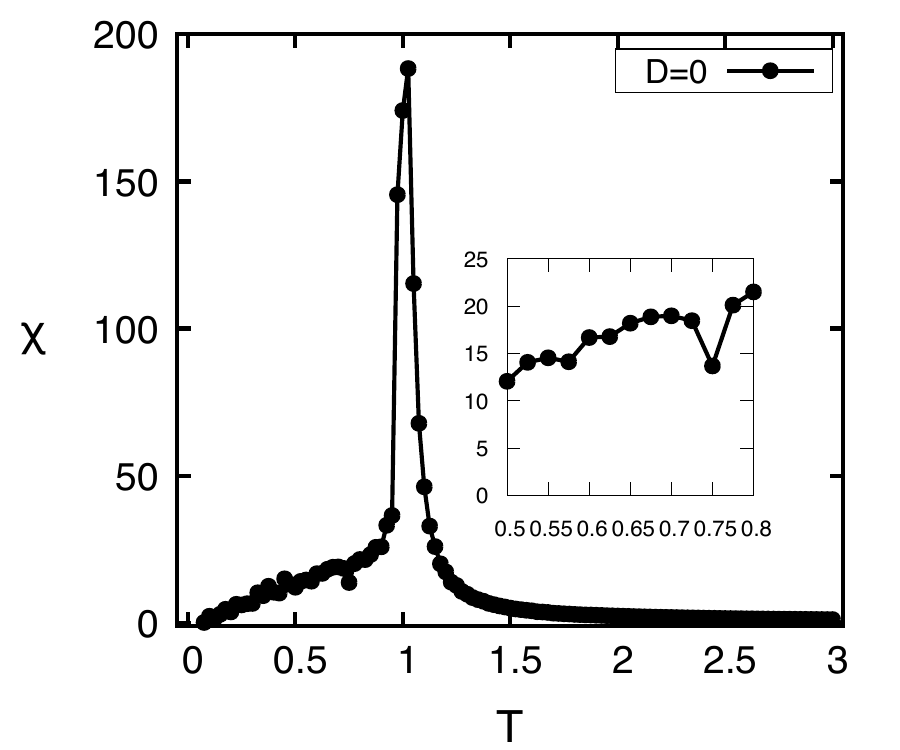}
    (c)\includegraphics[width=0.5\textwidth]{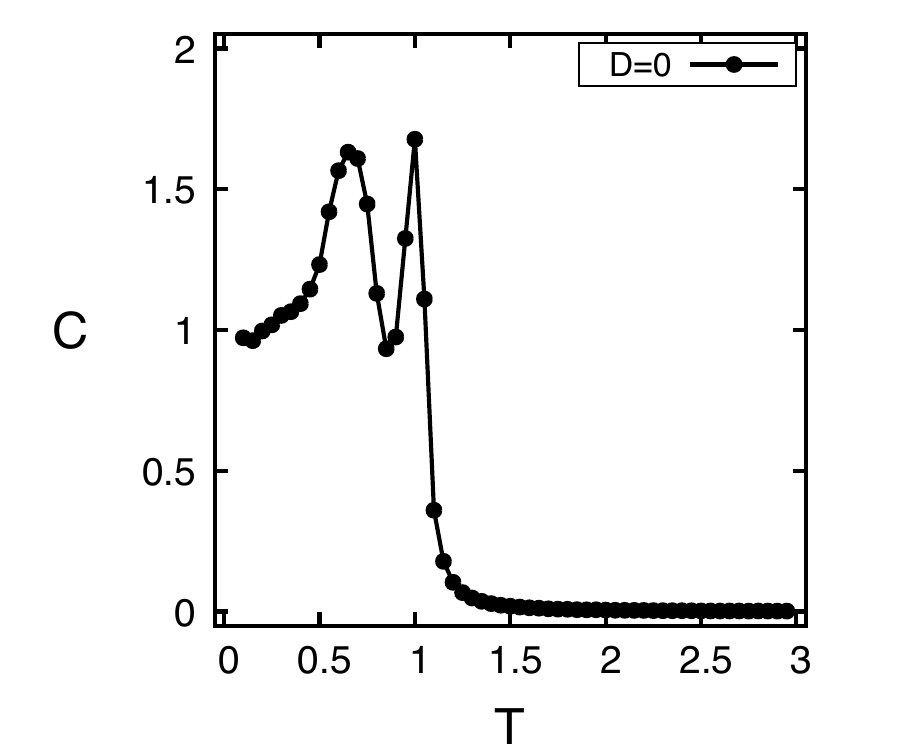}
	\caption{\label{D0}(a) The total magnetisation $(M)$ is plotted against temperature $(T)$ for isotropic $(D=0)$ $xy$ system. (b) The corresponding susceptibility $(\chi)$ is plotted against temperature $(T)$, inset: the enlarged view near the compensation point. (c) The specific heat $(C)$ is plotted against temperature $(T)$. }
		
\end{figure}

%\subsection{Isotropic system $(D=0)$}
We first consider the isotropic XY system ($D=0$), where the spins are free to orient in any direction within the interval $0 \leq \theta < 2\pi$. As the system is being cooled down from a high-temperature paramagnetic phase, the total magnetization 
starts to increase continuously at any finite temperature (critical temperature $T_c$), indicating the onset of long-range magnetic order, as shown in Fig-\ref{D0}(a). The corresponding magnetic susceptibility exhibits a pronounced peak at the critical temperature, $T_c$, identifying the  phase transition observed in Fig-\ref{D0}(b). Following further cooling below $T_c$, the total magnetization gradually decreases and eventually vanishes at a lower temperature, known as the {\it compensation temperature}, $T_{\mathrm{comp}}$. In the present trilayer ferrimagnetic system, this compensation arises from the competition between the temperature-dependent magnetizations of the three magnetic layers. Although each layer remains magnetically ordered (partially) below $T_c$, their unequal and a special combination of the values of the sublattice magnetisations lead to an exact cancellation of the net magnetization at $T_{\mathrm{comp}}$. As shown in Fig-\ref{D0}(c), the specific heat exhibits two characteristic indications: a sharp peak at the critical temperature and a second peak associated with the compensation temperature. In addition, a pronounced dip is observed in the vicinity of $T_{\mathrm{comp}}$ in the susceptibility curve as shown inset of Fig-\ref{D0}(b).

\begin{figure}[htbp]
     \centering
	(a)\includegraphics[width=0.5\textwidth]{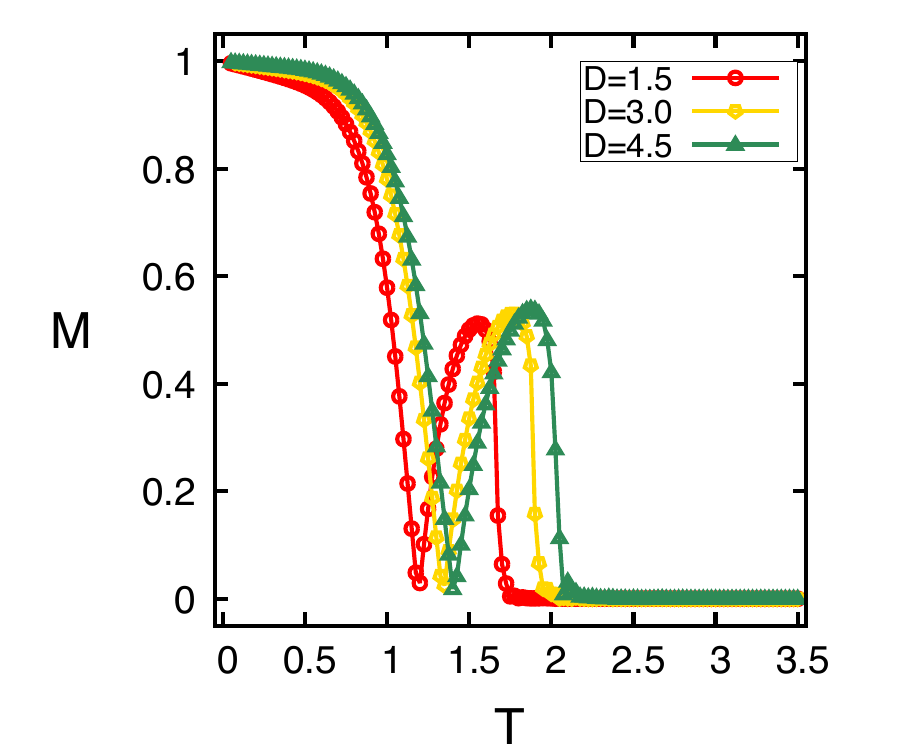}
	(b)\includegraphics[width=0.45\textwidth]{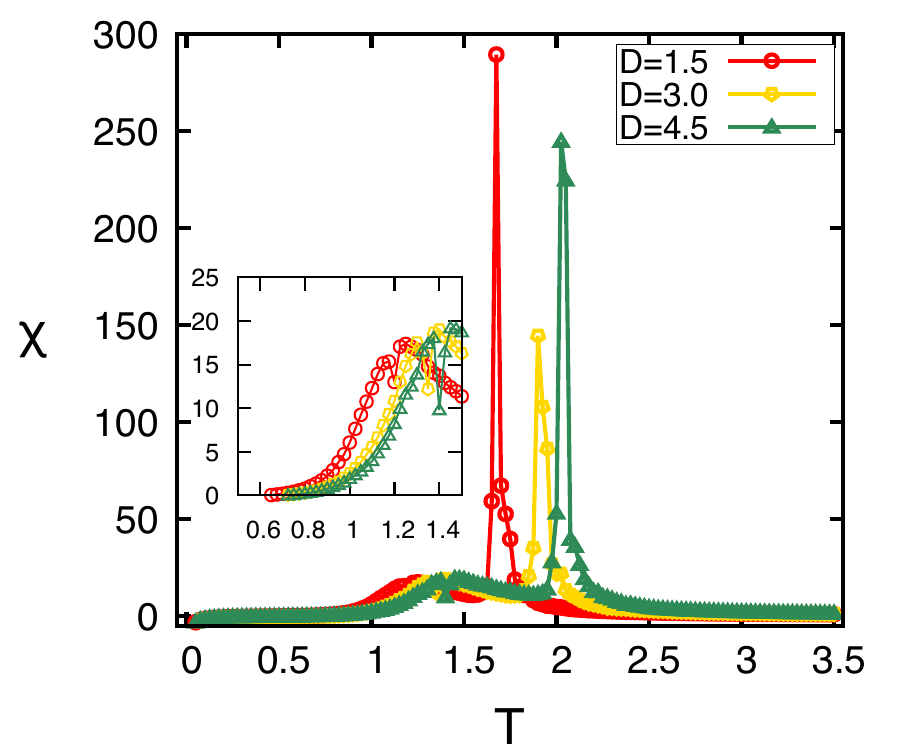}
    (c)\includegraphics[width=0.5\textwidth]{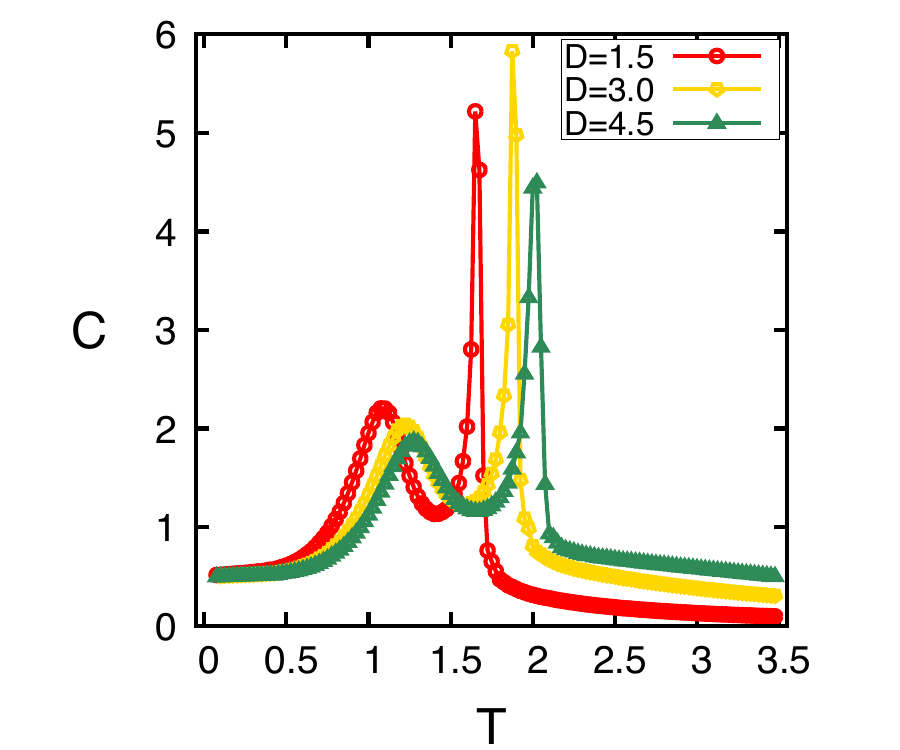}
	\caption{\label{D-observable}(a) The total magnetisation $(M)$ is plotted against temperature $(T)$ for different anistropy values $D=1.5, 3.0 ,4.5$. (b) The corresponding susceptibility $(\chi)$ is plotted against temperature $(T)$, inset: the enlarged view near the compensation point. (c) The specific heat is plotted against temperature $(T)$. }
		
\end{figure}

%%%%%%%%%%%%%%%%%%%%%%%%%%%%%%%%%%%%%%%%%%%%%%%%
We next study the effect of the single-site anisotropy, $D$, on the magnetic and thermodynamic properties of the trilayered XY system. Introducing a finite anisotropy breaks the continuous rotational symmetry by favouring spin alignment along a preferred direction, thereby reducing thermal spin fluctuations. Fig-\ref{D-observable}(a)-(c) present the temperature dependence of the total magnetization, magnetic susceptibility, and specific heat for several values of the anisotropy strength $D$. At high temperatures, the system remains in the paramagnetic phase, where thermal fluctuations dominate and the spins are randomly oriented, resulting in a vanishing total magnetization. As the temperature is lowered, the total magnetization increases continuously from zero, indicating the establishment of long-range ferrimagnetic order. The corresponding susceptibility exhibits a pronounced peak at the critical temperature, $T_c$, which accurately identifies the phase transition as illustrated in Fig-\ref{D-observable}(b). Upon further cooling, the total magnetization decreases and eventually becomes zero at the compensation temperature, $T_{\mathrm{comp}}$. In the trilayered system, this behaviour originates from the different kinds of temperature dependences of the sublattice magnetizations, whose opposing contributions exactly cancel each other while each layer remains magnetically ordered. Consequently, the system exhibits zero net magnetization without losing its sublattice magnetic order.

A systematic effect of the anisotropy is evident from the Figure-\ref{D-observable}. As $D$ increases, both the critical temperature and the compensation temperature shift toward higher values, demonstrating that the ordered ferrimagnetic phase becomes progressively more stable against thermal fluctuations. This enhancement arises because the anisotropy suppresses transverse spin fluctuations by energetically favoring alignment along the preferred direction, thereby requiring higher thermal energy to destroy the ordered state. The same trend is reflected in the specific heat curves, whose maxima move towards higher temperatures with increasing $D$. The simultaneous shift of the magnetization, susceptibility, and specific heat curves provides consistent evidence that the single-site anisotropy strengthens the magnetic ordering and extends its stability over a broader temperature range.

\begin{figure}[htbp]
    \centering
    (a)\includegraphics[width=0.5\textwidth]{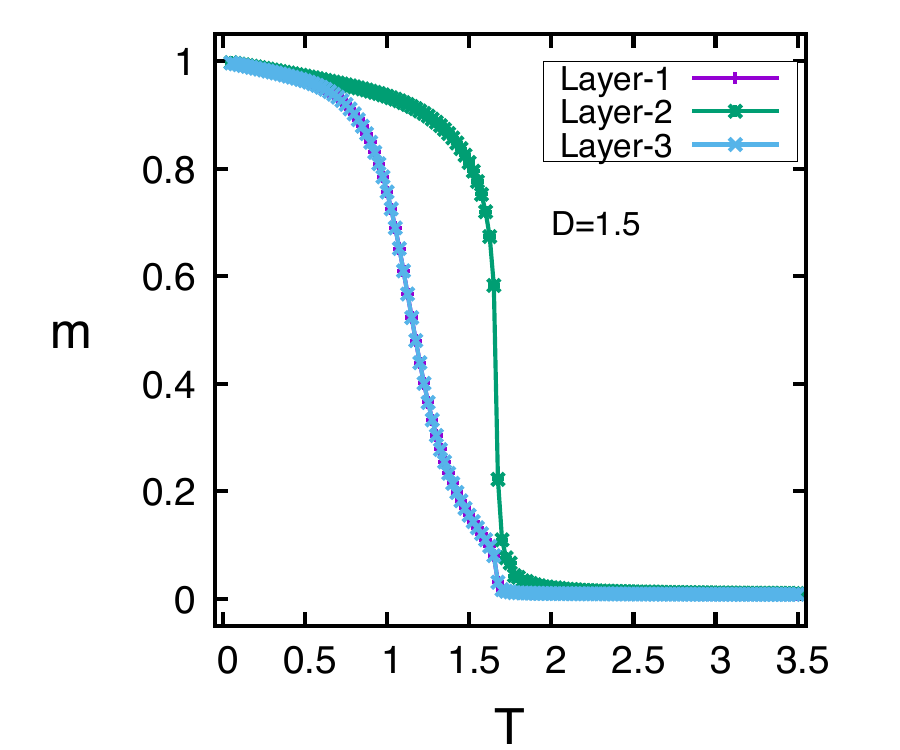}
    (b)\includegraphics[width=0.5\textwidth]{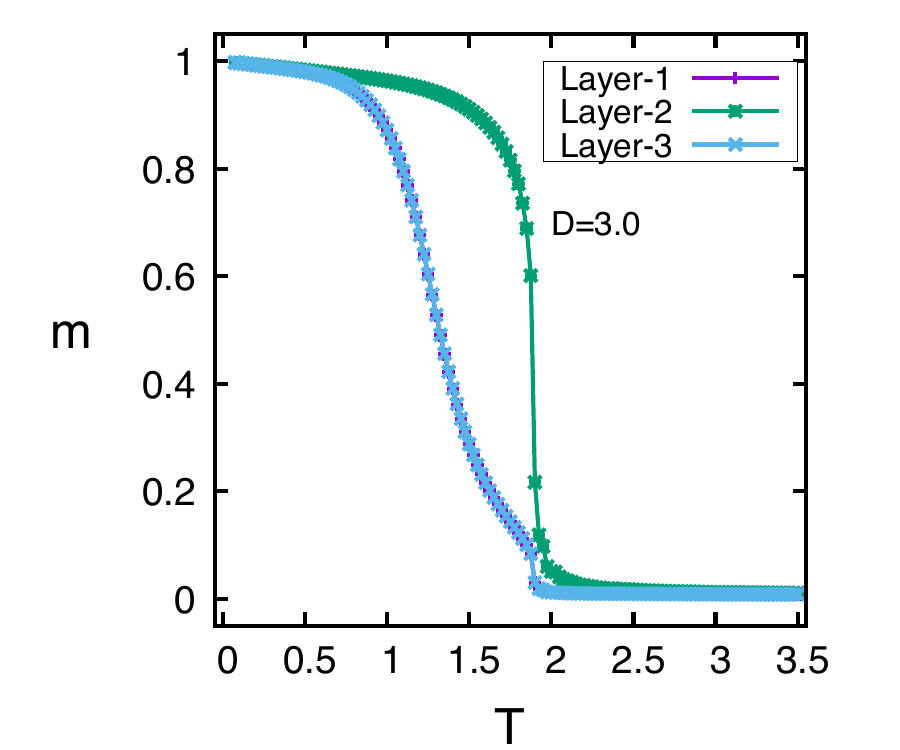}
    (c)\includegraphics[width=0.5\textwidth]{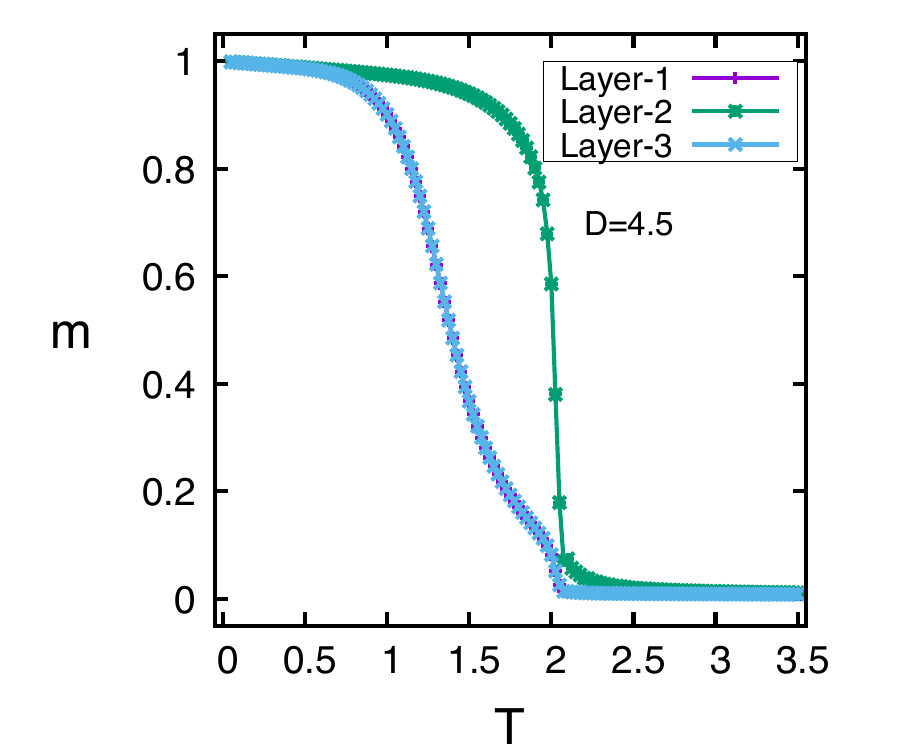}
	\caption{\label{sublattice-m} The sublattice (individual layer) magnetisation $m$ is plotted against temperature $(T)$ for (a)$D=1.5$ , (b)$D=3.0$ and (c)$D=4.5$. }
		
\end{figure}

To gain further insight into the origin of the compensation behaviour, we examine the temperature dependence of the sublattice magnetizations for the same values of the anisotropy strength $D$, as shown in Fig~\ref{sublattice-m}(a)-(c). Owing to ferromagnetic intralayer interactions, a fraction of the  spins within each layer align parallel below the critical temperature. This creates a finite spontaneous magnetization within the layers, even if long-range ferromagnetic order is lost globally.In contrast, the antiferromagnetic interlayer couplings favour antiparallel alignment between adjacent layers, giving rise to a ferrimagnetic ordering.

 As the temperature is lowered further, the magnitudes of the sublattice magnetizations vary at different rates. This is due to different ferromagnetic intraplanar interactions. At the compensation temperature, the magnetization of one layer (B) is exactly balanced by the combined magnetizations of the other two layers (A), causing the total magnetization to vanish while each  layer remains magnetically ordered. This clearly demonstrates that the compensation phenomenon is not merely associated with the disappearance of magnetic order (due to random spin alignments) but rather results from the exact cancellation of the oppositely aligned sublattice magnetizations.

\begin{figure}[htbp]
   \centering
	(a)\includegraphics[width=0.6\textwidth]{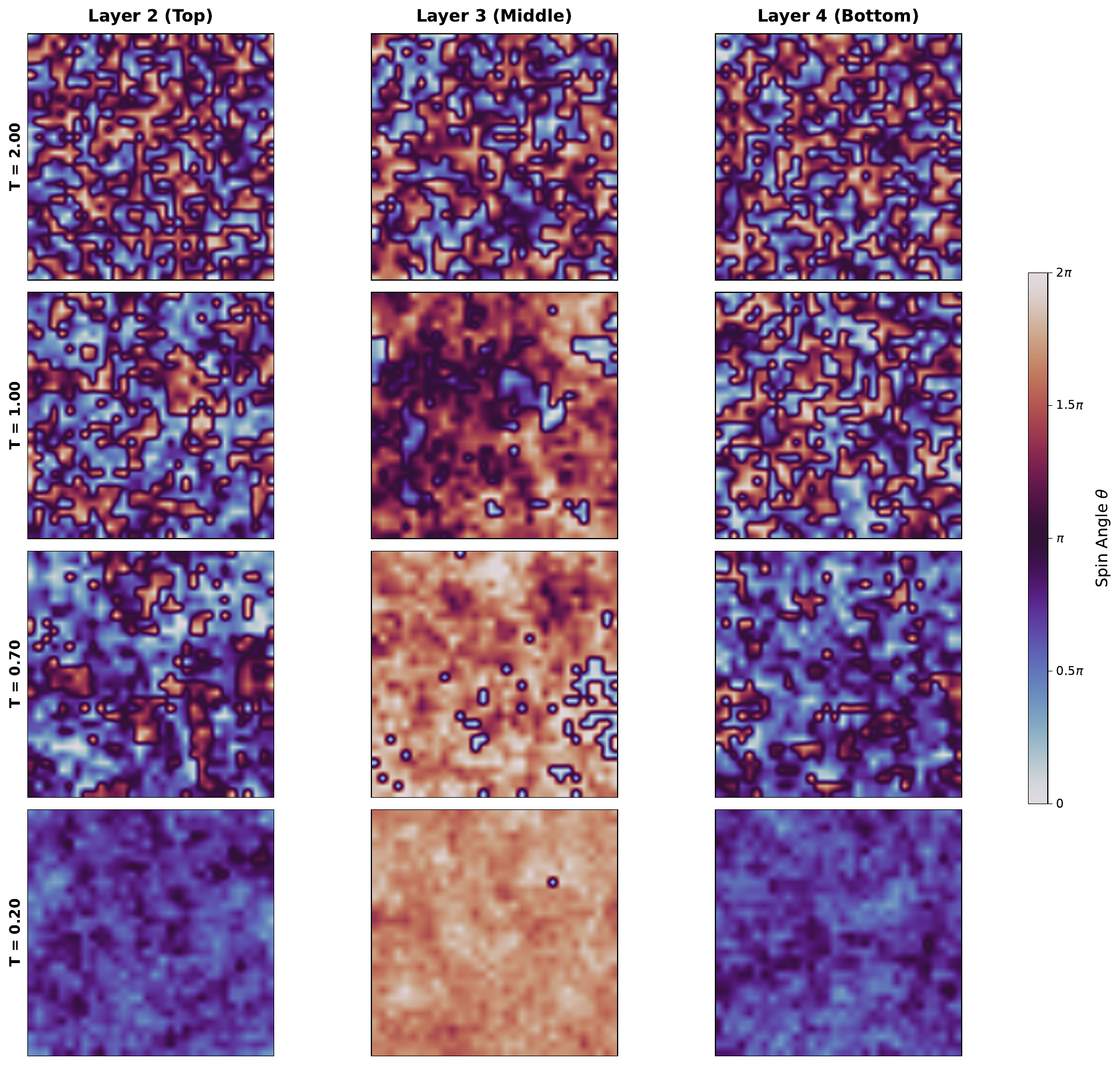}
    (b)\includegraphics[width=0.6\textwidth]{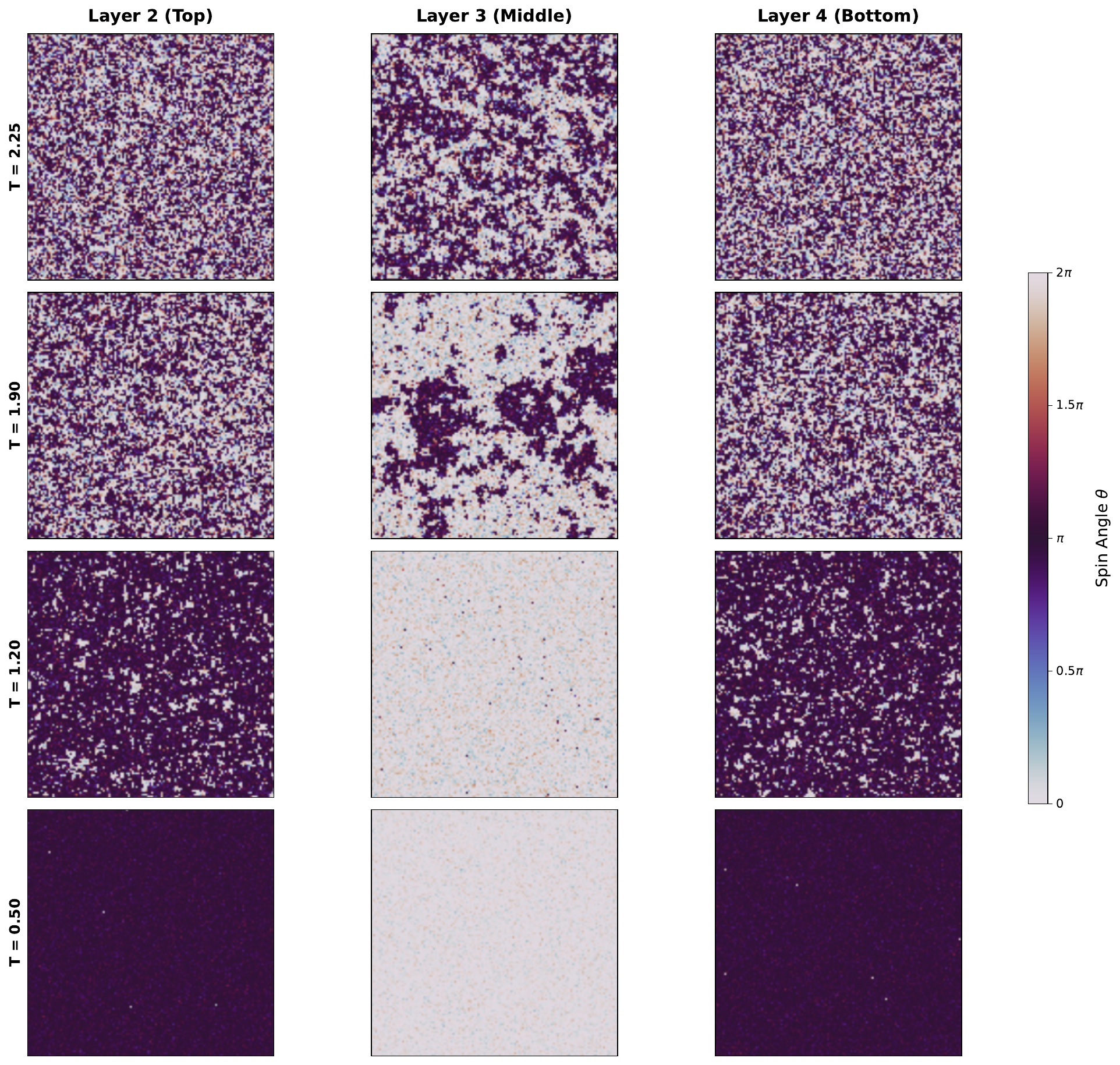}
	\caption{\label{spin-morphology}Spin morphology for system size $L=32$ at (a)isotropic regime $(D=0)$ and (b) anisotropic regime $(D=3)$.  For each regime, spin configurations are shown for the three layers at four temperatures: high temperature, near critical temperature, near compensation temperature and low temperature, respectively. The orientations of spins represent their corresponding spin angles in the xy plane. The figures illustrate the evolution of spatial spin arrangement with temperature and highlight the differences in spin morphology between isotropic and anisotropic regimes. }
		
\end{figure}

%%%%%%%%%%%%%%%%%%%%%%%%%%%%%%%%%%%%
%\subsection{Spin Morphology}
To gain microscopic insight into the magnetic ordering process, we observe the spin morphologies of the individual layers at four representative temperatures corresponding to the paramagnetic, critical, compensation, and low-temperature regimes. The spin configurations are represented by heat maps in which each color denotes a particular spin orientation (angles). Fig-\ref{spin-morphology}(a) show the isotropic system ($D=0$), whereas Fig-\ref{spin-morphology}(b) presents the corresponding configurations for single-site anisotropy $D=3.0$. The evolution of the spin morphologies reflects the competition between the ferromagnetic intralayer exchange interactions, which promote parallel alignment of spins within each layer, and the antiferromagnetic interlayer coupling, which favors antiparallel alignment between neighbouring layers. For the isotropic system, all spin orientations in the interval $0 \leq \theta < 2\pi$ are energetically favourable. Consequently, at high temperatures the spin-angle distribution is nearly uniform, producing numerous small domains with widely varying spin orientations in each layer, characteristic of the paramagnetic phase. As the temperature approaches the critical temperature, thermal fluctuations are progressively overcome by the exchange interactions, leading to domain coarsening and the emergence of large correlated regions with similar spin orientations. Below the critical temperature, each layer develops long-range magnetic order. At the compensation temperature, the two outer layers and the middle layer form large, well-defined domains with opposite dominant spin orientations due to the antiferromagnetic interlayer interaction. Because the sublattice magnetizations evolve differently with temperature, the magnetization of the middle layer exactly balances the combined magnetization of the two outer layers, resulting in zero total magnetization while each layer remains magnetically ordered. Upon further cooling, the system approaches the antiferromagnetic ground state, where each layer becomes nearly single-domain and the antiparallel alignment (between the outer and middle layers) is clearly established.

\begin{figure}[htbp]
   \centering
	(a)\includegraphics[width=0.45\textwidth]{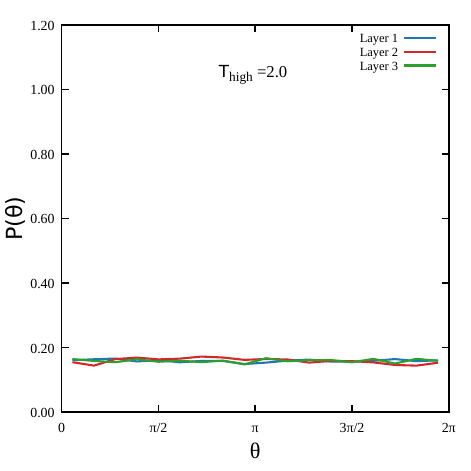}
	(b)\includegraphics[width=0.45\textwidth]{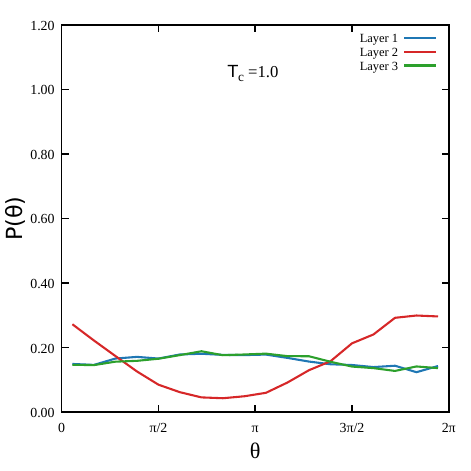}
        (c)\includegraphics[width=0.45\textwidth]{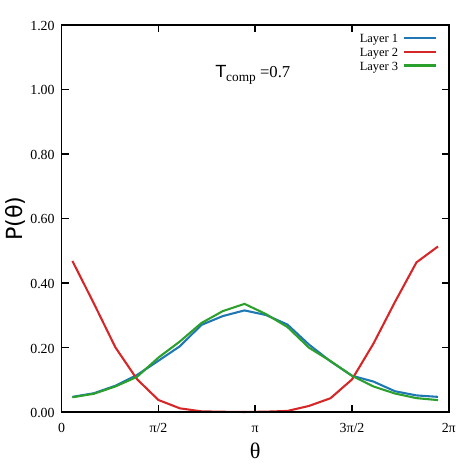}
        (d)\includegraphics[width=0.45\textwidth]{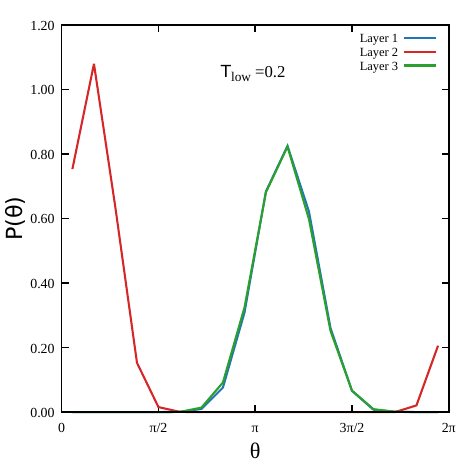}
	\caption{\label{spin-density} The spin-angle probability distribution  of three sublattices for four characteristic temperatures for isotropic XY system. }
		
\end{figure}

The spin morphology changes significantly in the presence of single-site anisotropy as illustrated in Fig-\ref{spin-morphology}(b). Since the favourable alignment is along the $x$ direction, the anisotropy energetically favours spin orientations around $\theta=0$ and $\theta=\pi$, strongly suppressing transverse 
spin fluctuations. Even in the high-temperature paramagnetic phase, most spins fluctuate around these two preferred orientations, with only a small fraction occupying intermediate angles due to thermal excitations. The heat maps therefore exhibit a much narrower distribution of spin orientations than in the isotropic case, while both the individual layer magnetizations and the total magnetization remain approximately zero. Near the critical temperature, the ferromagnetic intralayer interactions establish long-range order within each layer, whereas the antiferromagnetic interlayer coupling aligns the middle layer antiparallel to the two outer layers. Stronger ferromagnetic coupling acts in the middle layer (B). Consequently, the outer layers are predominantly oriented around $\theta=\pi$, with only small regions near $\theta=0$, while the middle layer is predominantly oriented around $\theta=0$ with a few spins near $\theta=\pi$. At the compensation temperature, this antiparallel arrangement becomes even more pronounced: the individual layer magnetizations remain finite, yet the magnetization of the middle layer exactly compensates that of the two outer layers, producing zero net magnetization. At low temperatures, thermal fluctuations get significantly reduced, and each layer evolves into an almost perfect single-domain state, with the two outer layers aligned along $\theta=\pi$ and the middle layer along $\theta=0$ (or equivalently $2\pi$). To quantify spin morphology statistically, we compute the spin-angle probability distribution, $P(\theta)$, for the isotropic system. The statistical distribution of the spin orientation (angles) (shown in Fig-\ref{spin-density}) captures these features. 

%%%spinangledensity%%%%%%%%%%%%

%%%%%%%%%%%%%%%phasediagram%%%%%%%%%%%%%%%%

\begin{figure}[htbp]
   \centering
	\includegraphics[width=0.8\textwidth]{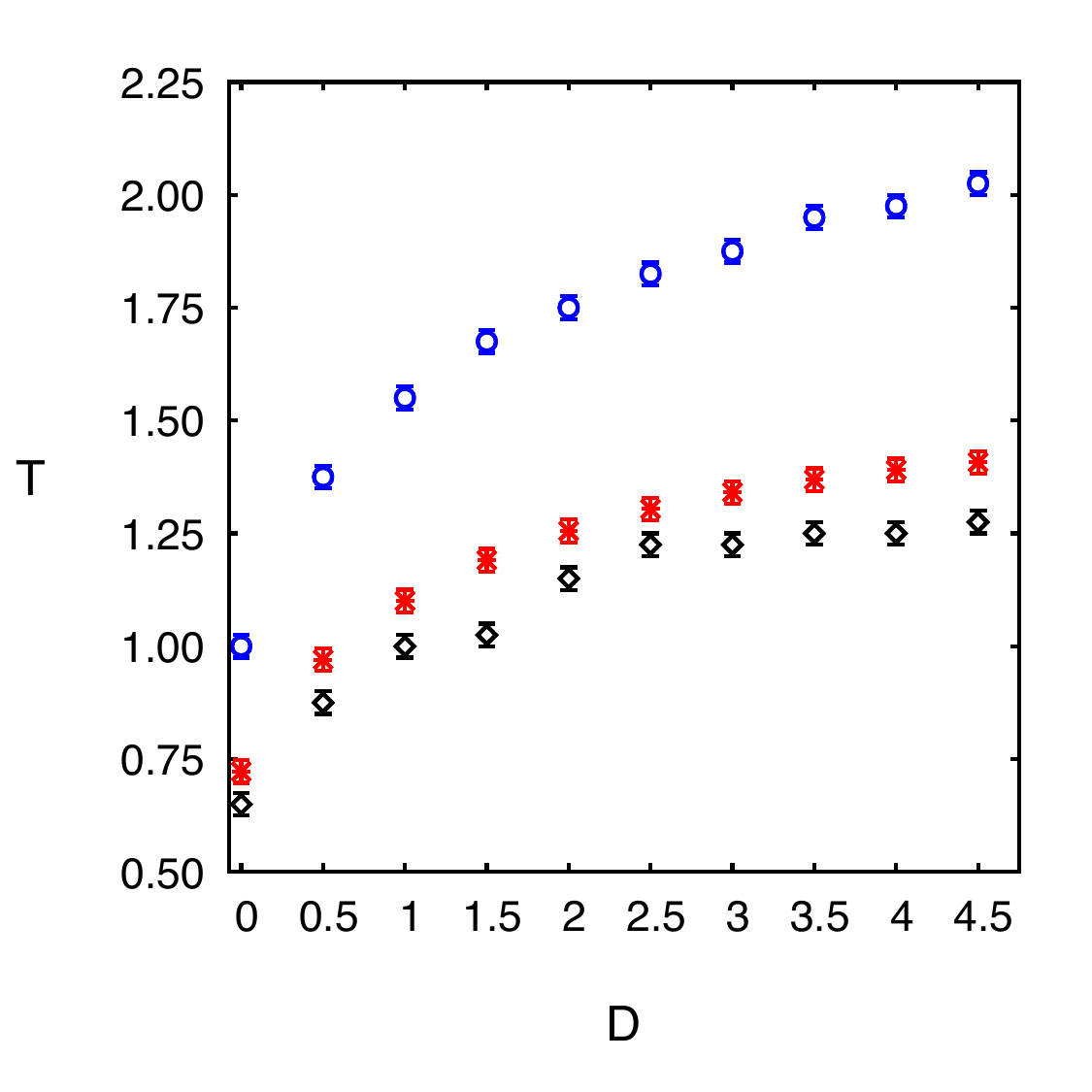}
	\caption{\label{phase} Phase diagram in $T-D$ plane. The critical temperature $T_c$ is represented by \textit{blue} $\circ$ while the compensation temperature $T_{comp}$ is denoted by \textit{black} $\diamond$. The interpolated compenstation temperature $T_{comp}^{int}$  shown by \textit{red} $\star$, is obtained from interpolation between two successsive temperatures at which magnetization changes sign. The error bars corresponds to an uncertainty of $\pm0.025$, arising from the finite tempearture interval used to locate the transition or compensation point. }
\end{figure}

\begin{figure}[htbp]
   \centering
	\includegraphics[width=0.8\textwidth]{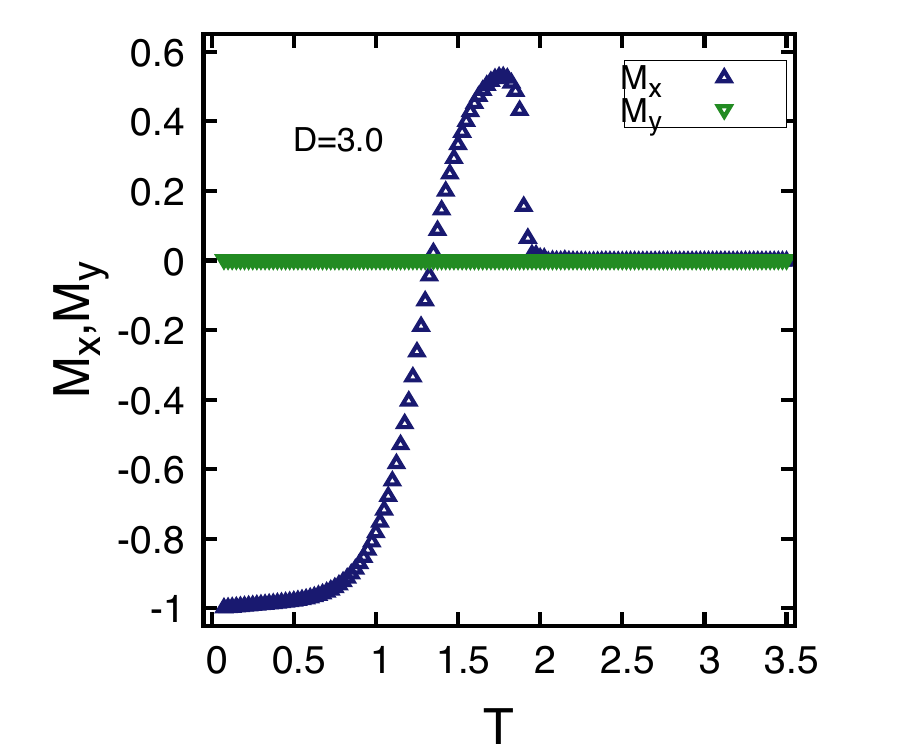}
	\caption{\label{MxT} Temperature dependences of total magnetisation components $M_x$ and $M_y$ for $D=3.0$. The vanishing of $M_x$ at lower temperature indicates the compensation, whereas
	$M_y$=0 for all $T$. }
\end{figure}

The overall phase behaviour of the trilayer XY system is summarized in the $D$--$T$ phase diagram shown in Fig-\ref{phase}. The phase boundaries are obtained by determining the critical temperature, $T_c$, from the first peak (higher temperature) and the compensation temperature, $T_{\mathrm{comp}}$, from the low-temperature peak in the specific heat. The phase diagram clearly demonstrates that both the critical and compensation temperatures increase monotonically with increasing single-site anisotropy. This behaviour indicates that the anisotropy restrains thermal spin fluctuations and enhances the stability of the ferrimagnetic phase. However, the increase in $T_{\mathrm{comp}}$ is noticeably weaker than that of $T_c$, implying that the anisotropy has a stronger influence on the establishment of long-range magnetic order than on the balance between the oppositely aligned layer magnetizations responsible for the compensation phenomenon. To verify the determination of the compensation temperature, we employ an independent procedure based on the temperature dependence of the total magnetization. In the anisotropic system, the favourable direction is the $x$ direction, and consequently the transverse magnetization remains zero ($M_y=0$), while the longitudinal component $M_x$ changes sign in the vicinity of the compensation point (shown in Fig-\ref{MxT}). The sign reversal of $M_x$ reflects the inversion of the net magnetic moment arising from the unequal thermal evolution of the oppositely aligned sublattice magnetizations. The temperatures immediately above and below the sign change are identified, and linear interpolation between the corresponding magnetization values is used to determine the temperature at which the total magnetization component ($M_x$) vanishes.
The compensation temperatures obtained from this interpolation procedure are represented by the red symbols in Fig-\ref{phase}. These values are slightly above of those computed from the low temperature maxima of the specific heat. 

\begin{figure}[htbp]
	(a)\includegraphics[width=0.5\textwidth]{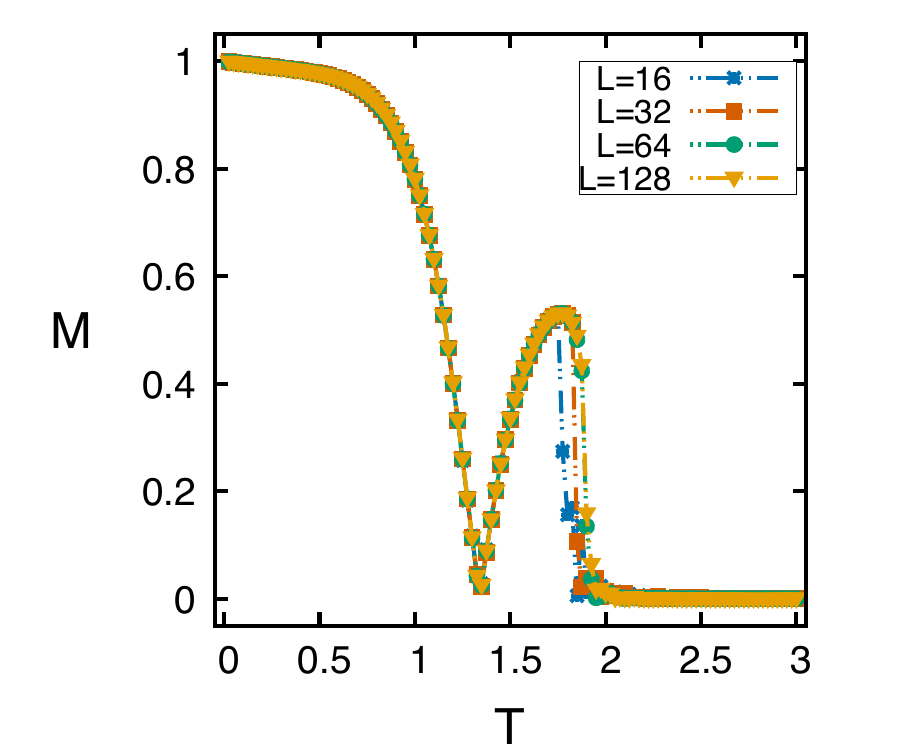}
    (b)\includegraphics[width=0.45\textwidth]{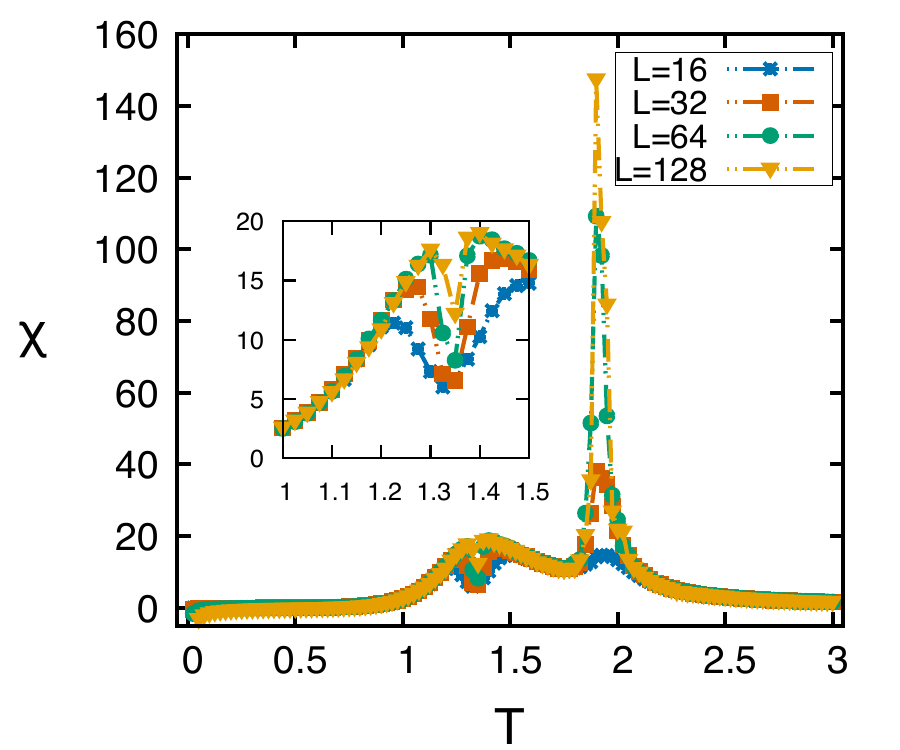}
    (c)\includegraphics[width=0.5\textwidth]{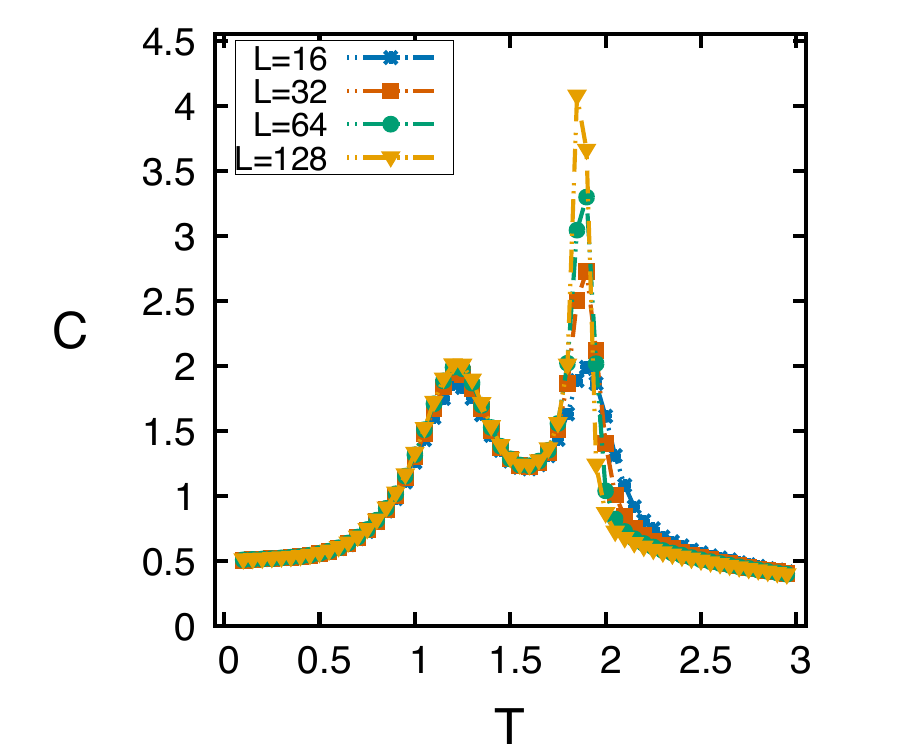}
	\caption{\label{Finite-size} The thermal variation of (a) the magnetization (b) the susceptibility and (c) the specific heat for different system sizes $L$ at anisotropy strength $D=3.0$. The systematic variation with $L$, particularly in the vicinity of transition illustrate the finite-size dependence of the observed thermal behaviour. The inset of (b) shows the enlarged view of susceptibility curve near the compensation temperature.  }
		
\end{figure}

%%%Finite size analysis %%%%%%%%%
The presence of a thermodynamic phase transition and the associated development of long-range order have been examined through finite-size study. Near a continuous phase transition, the correlation length grows rapidly as the transition temperature is approached, leading to pronounced size-dependent behaviour of the susceptibility. In particular, the susceptibility is expected to exhibit an increasing height of the peak with increasing system size as the system approaches the thermodynamic limit. To examine the nature of the observed magnetic transition, we have studied the thermal variations of the magnetization, susceptibility, and specific heat for different system sizes $L$, as shown in Fig.~\ref{Finite-size}. As evident from Fig.~\ref{Finite-size}(b), the maximum value of the susceptibility associated with the critical temperature $T_c$ increases systematically with increasing system size. A similar size dependence is observed in the specific heat [Fig.~\ref{Finite-size}(c)], where the peak at $T_c$ becomes more pronounced as $L$ increases. This systematic growth of the response-function peaks with system size is consistent with the growth of critical correlations near the transition and supports the existence of a genuine thermodynamic phase transition, separating the long-range ordered phase below $T_c$ from the disordered phase above it.

In contrast, the low-temperature features associated with the compensation temperature do not exhibit similar critical behaviour. In particular, no significant size dependence is observed in the low-temperature peak of the specific heat. Moreover, the dip in the susceptibility curve in the vicinity of the compensation temperature becomes progressively weaker as the system size increases, as shown in the inset of Fig-\ref{Finite-size}(b) and eventually disappears for sufficiently large $L$. This clearly indicates that the existence of the dip embedded in the low temperature smeared peak of the susceptibility is clearly a finite-size effect.  The compensation phenomenon arises from the cancellation of the oppositely oriented sublattice magnetizations, resulting in a vanishing total magnetization while the individual sublattice magnetisations remain nonvanishing. Therefore, unlike the transition at $T_c$, the compensation temperature does not correspond to a conventional thermodynamic phase transition. 

\section{Summary:}
\label{sec:conclusions}
The trilayered anisotropic XY (SO(2) continuous symmetric) model has been considered with intraplanar ferromagnetic and interplanar antiferromagnetic interactions. The system has been studied extensively by Monte Carlo simulation with Metrolpolis random single-spin update algorithm.
The sublattice magnetisation, total magnetisation, magnetic susceptibility and the specific heat are studied as function of the temperature of the system. Upon cooling from high temperature paramagnetic phase, the system was found to exhibit phase transition. Below the transition temperature an
interesting phenomenon occurs, where the total magnetisation vanishes even for nonzero sublattice magnetisations. This is a very special combination of sublattice magnetisation (with suitably chosen interaction parameters) to make the total magnetisation zero. This is called {\it compensation}. We have studied the compensation temperature and the critical temperature as functions of the anisotropy of the system. In the vicity of the critical and compensation temperature the spin configurations are described by the distribution of angles. The susceptibility and the specific heat show peak near the compensation and critical temperature. The comprehensive phase diagrams are drawn. The critical and compensation temperatures are found to increase with the increase of single site anisotropy of the system. As far as the knowledge of the authors are concerned, the present study is the first in the literature to report the Monte Carlo results of compensation behaviours in continuous symmetric spin models.

Some interesting effects may be studied further in planar ferrimagnet. The effects of the compensation behaviour in XY ferrimagnet for random disorder i.e.,
random field\cite{xy-rf}, random anisotropy\cite{xy-anis} and random impurity\cite{xy-imp}. The effects of random bonds (ferromagnetic/antiferromagnetic) may also be an interesting study. Moreover, the anisotropic Heisenberg model may be a good choice to investigate such compensation behaviours in continuous symmetric spin models.

\vskip 0.5cm

\noindent {\bf Acknowledgements:} We thankfully acknowledge the discussion with Anusrita Mukherjee. The computational facility provided by Presidency University is gratefully acknowledged.

\vskip 1cm
\noindent {\bf Data availability statement:} Data will be available on reasonable request to Olivia Mallick.

\vskip 0.2cm

\noindent {\bf Conflict of interest statement:} We declare that this manuscript is free from any conflict of interest.
\vskip 0.2cm

\noindent {\bf Funding statement:} No funding was received, particularly to support this work.

\vskip 0.2cm

\noindent {\bf Authors’ contributions:} Olivia Mallick developed the code, collected the data, prepared the figures, analysed the results and wrote the manuscript.
Muktish Acharyya conceptualised the problem, analysed the results and wrote the manuscript.

%%%%%%%% FIGURES Main Text %%%%%%%%

\newpage 

%%%%%%%%%%%%%%%%%%%Figure-1%%%%%%%%%%%%%%%%%%%%%%%%%%%%%%%%%%%%%%%%%%%%%%%%%%%%
%%%%%%%%%%%%%%%%%%%%%%%%%%%%%%%%%%%%%%%%%%%%%%%%%%%%%%%%%%%%%%%%%%%%

\newpage

%%%%%%%%%%%%%%%%%%%%%%%%%%%%%%%%%%%%%%%%%%%%%%%%%%
%%%%%%%%%%%%%%%%%%Figure-2%%%%%%%%%%%%%%%%%%%%%%%%%%%%%%%%%
%%%%%%%%%%%%%%%%%%%%%%%%%%%
%%%%%%%%%%%%%%%%%%%%%%%%%%%%%%%%%%%%%%%%%%%%%%%%%%%%

\newpage
%%%%%%%%%%%%%%%%%%%%%%%%%%%%%%%%%%%%%%%%%%%%%%%%%%%%%%%%
%%%%%%%%%%%%%%%%%%Figure-3%%%%%%%%%%%%%%%%%%%%%%%%%%%%%%%%%
%%%%%%%%%%%%%%%%%%%%%%%%%%%

\newpage
%%%%%%%%%%%%%% FIGURE-4%%%%%%%%%%%%%%%%%%%%%%%%%%

%%%%%%%%%%%%%%%%%%%%%%%%%%%%%%%%%%%%%%%%%%%

\newpage
%%%%%%%%%%%% FIGURE-5 %%%%%%%%%%%%%%%%%%%%%%%%%%
%%%%%%%%%%%%%%%%%%%%%%%%%%%%%%%%%%%%%%%%%%%%%%%%%%%%

\newpage
%%%%%%%%% FIGURE-6 %%%%%%%
%%%%%%%%%%%%%%%%

\newpage
%%%%%%%%% FIGURE-7 %%%%%%%
%%%%%%%%%%%%%%%%

%%%%%%%%%%%%%%%%%%% FIGURE-8%%%%%%%%%%%%%%%%
\newpage

%\begin{figure}[htbp]
	%\includegraphics[width=79mm]{}
	%\includegraphics[width=80mm]{}
	%\caption{\label{}}
%\end{figure}

\newpage

\end{document}